\documentclass[a4paper, 11pt, twoside]{article}
\usepackage{CJK}
\usepackage{latexsym,bm}
\usepackage[tbtags]{amsmath}
\usepackage{amssymb}

\usepackage{fancyhdr}
\usepackage{cite}
\usepackage{tabularx}
\usepackage{booktabs}
\usepackage{dcolumn}

\usepackage{multirow}

\usepackage{graphicx}

\usepackage{wrapfig}
\usepackage{multicol}
\usepackage{verbatim}

\def\nkyear{0}                           
\def\nkvol{0}                              
\def\nkno{0}                                
\def\firstpage{1}                           
\def\DOI{0}
\def\shorttitle{Aerodynamics and mechanisms of
morphing wing}  
\def\shortauthor{Z.W. GUAN \& Y.L. YU}
\usepackage{Shangda}                    

\newcommand{\supercite}[1]{\!\!\textsuperscript{\cite{#1}}} 

\begin{document} 

\begin{CJK*}{GBK}{Song}

\title{
{\large \textbf{A study on aerodynamics and mechanisms of
elementary morphing models for flapping wing
in bat forward flight}
}
\thanks{
\footnotesize{Received Dec.\;xx, 20xx / Revised Jan.\;xx, 20xx
\newline  
Project supported by the National Natural Science Foundation of China (No.\,10602061)
\newline
Corresponding YU Yong-Liang, Ass. Prof., Ph.\,D.,
E-mail:\;ylyu@ucas.ac.cn, ylyu@ustc.edu}}
} 

\author{
\small{GUAN Zi-Wu$^{1,2}$, \quad YU Yong-Liang$^1$} 
\\[2mm]
\footnotesize{(1. The Laboratory for Biomechanics of Animal Locomotion,
University of Chinese Academy of} \\
\footnotesize{Sciences, Beijing 100049, China;} 
\\
\footnotesize{2. Department of modern Mechanics,
University of Science and Technology of China, Hefei 230027,}\\
 \footnotesize{China)} 
\\
}

\maketitle 
\thispagestyle{first}
\footnotesize
\begin{abstract}
\noindent \textbf{Abstract~~~} The large active wing deformation is a significant way to generate high aerodynamic forces required in bat flapping flight. Besides the twisting, the elementary morphing models of a bat wing are proposed, such as wing-bending in the spanwise direction, wing-cambering in the chordwise direction, and wing area-changing. A plate of aspect ratio $3$ is used to model a bat wing and a three dimensional unsteady panel method is applied to predict the aerodynamic forces. It is found that the cambering model has a great positive influence on the lift, followed by area-changing model and then the bending model. The further study indicates that the vortex control is a main mechanism to produce high aerodynamic forces, and the mechanisms for the aerodynamic force enhancement are the asymmetry of the cambered wing and the amplifier effects of wing area-changing and wing bending. The lift and thrust are mainly generated during the downstroke and almost negligible forces during the upstroke by the integrated morphing model-wing.
\\[2mm]
\textbf{Key words~~~}bat wing, bending, cambering, area-changing, aerodynamic force
\\[2mm]
\textbf{Chinese Library Classification~~~} O351.2, O355 \\
\textbf{2010 Mathematics Subject Classification~~~}76B47, 76M23, 76Z10
\end{abstract}

\section{Introduction} 

The bat flight, especially the moderate or small bat flight, attracts more and more researchers to study the aerodynamic mechanisms of the flexible wings in the last decades. In the order of their size (about 10cm), the fixed bat wing cannot generate the sufficient aerodynamic forces to sustain the flight at low speeds (less than 5m/s), since there is no enough lift to balance the weight and no thrust generated. Therefore, the aerodynamics of actively deforming bat-wing has been paid much more attention and it takes us a lot of attractive imaginations. For the scientists and engineers who work on micro air vehicles (MAVs), bats provide an important inspiration to bionic designs.

The large active wing-deformation is widely observed and it may be an effective way to produce the aerodynamic forces required in free flight. On the one hand, bat is the only flying mammal capable of powered flight, who requires higher lift than birds or insects in its forward flight because of their larger mass density. On the other hand, its unique integrated wing-body structure limits the wing flapping like birds or insects. It is known that a bat wing is formed by elastic muscularized membranes and an upper limb. The flexible membranes are stretched between the digits of the hand, the hindlimb and the body wall, and make up the lifting surface of the wing. The upper limb, similar to a human's, with the wing muscles plays a key role in controlling the joint motions, such as the flexion/extension of the elbow or the digits, and/or ab/adduction of the wrist, which could cause changes in camber, bending or twisting of the wing\supercite{1}. The distinct arrangement of the bat wing suggests a high potential for ability to adjust the wing morphing according to the aerodynamic demands in flight\supercite{2}. However, for the complicated deformations of bat flapping wing, some questions are waiting to be answered. For example, how to model the actively deforming bat-wing in flapping flight, and how the actively deforming affects the aerodynamic forces acting on the bat wing during flight?

In our previous study\supercite{3}, the deformation called `twisting' which was defined as the AOAs varying along the spanwise direction was modeled and investigated. The results showed that a bat could obtain thrust by twisting its wings, and the twisting in bat flight has the same function as the supination/pronation motion in insect flight\supercite{3}. Besides the twisting, wing morphing such as bending, cambering and wing-area changing are also observed during bat flapping flight.

In the last decades, a lot of experimental works have been done in studying bat flight, which provide sufficient data to model the complicated active deformation. The kinematics, which included flapping frequency, strokeplane angles and amplitudes, etc., and ecological morphology were presented by Norerg\supercite{4,5}. In 1986, Aldridge found that wingbeat frequency decreased and wingbeat strokeplane angles increased with flight speed increasing\supercite{6}. The Swartz group did a systematical research in bat flight. They revealed that, at relatively slow flight speeds, the wing motion was quite complex, including a sharp retraction of the wing during the upstroke and a broad sweep of the partially extended wing during the downstroke\supercite{7}. They also quantified the complexity of bat wing kinematics\supercite{8}. Recently, Busse et al.\supercite{9} gave details of three dimensional wingbeat kinematics including wing movement, frequency, stroke plane angles, wing camber in the section of the fifth digit, angles of attack (AOAs), etc. The steady-state aerodynamic and momentum theories were used to predict the aerodynamic forces in hovering flight in the early days\supercite{4,5,10}. The wakes were captured to estimate the forces in different flight speeds$^{[11\textrm{--}14]}$.

In the present paper, we'll model the elementary deformations of bat-wing bending, cambering and area-changing, then use a model-wing to investigate their aerodynamic forces with the previously developed panel method which has been proved to be efficient and sufficiently credible in the study on the aerodynamic forces acting on the twisting model-wing in flapping\supercite{3}.

\section{Elementary morphing models}
There are two kinds of kinematic variables to describe the motion of bat wing, one is related to the flapping motion (i.e. frequency, amplitude and stroke plane), the other is related to the morphing motion (i.e. twisting, cambering, bending and wing area-changing). In the previous study\supercite{3}, the flapping and the twisting models have been studied according to the experimental data published by Busse et al.\supercite{9}, where the mean chord length of the bat wing is about 4.7cm, the aspect ratio (one wing) is about 3 and the frequency is 11.13Hz when it flies at a speed of 3m/s. Here, the mean chord length, the forward speed and the density of the air are selected as the references to nondimensionalize all of the variables.

In the present work, we have to investigate the aerodynamic forces caused by every elementary morphing motion. A rectangular plate with the aspect ratio of 3 is used to model the bat wing. Thus, the aerodynamic forces of a complex geometrical shape will be studied in future on the basis of these results of several elementary morphing models.

\subsection{The flapping model and twisting model}
Details of the flapping model and twisting model have been described in the paper\cite{3}, so only the characteristics are presented here. Flapping is a common motion in animal powered flight and it is assumed naturally that the flapping with respect to angular movement is sinusoidal in bat flight. So the flapping angle is
\begin{equation}
\theta(t)=\theta_0+\theta_A\cos \omega t
\end{equation}
where $\theta_0$ is the average flapping angle,
$\theta_A$ is the flapping amplitude, $\omega$ is the angular frequency and $t$ is the time that are nondimensionlized. In the paper, $\theta_A = 40\,^{\circ}$ and $\omega = 1$. For the symmetric flapping, $\theta_0$  is set to be zero, and the moment when the wing tip is located in the upper-most position is set to be the initial time.

The twisting model represents the distributions of AOAs along spanwise direction. With the linear distribution hypothesis, two parameters are used to assess the twist-morphing, which are AOA at wing root ($\alpha_{root}$) and AOA at wing tip ($\alpha_{tip}$). $\alpha_{root}$ is a constant about $7\,^{\circ}$, while $\alpha_{tip}$ is
\begin{equation}
\alpha_{tip}=\alpha_{tip0}+\alpha_A\cos(\omega t + \varphi_{t})
\label{Eq:alpha_tip}
\end{equation}
where $\alpha_{tip0}$ is the average value and $\alpha_A$ is the twisting amplitude. Here, $\alpha_{tip0}=15\,^{\circ}$ and $\varphi_{t} = \pi/2$. If the purely flapping motion is considered, $\alpha_A$ is set to be zero and the wing is rigid.

\subsection{Morphing models}
To model the actively deforming of bat wing, some hypotheses have to be brought in. Similar to the twisting model, the harmonic and rhythmic motion is introduced in the following morphing models, i.e. bending, cambering and area-changing, where the dynamic shape of the wing is assumed to be an arc in the bending and cambering models.

\subsubsection{Bending model}
\begin{figure}[h]
\centering \mbox{
\subfigure[A bat in flight (from Song\supercite{15}). The wing shape looks like an arc.]{
\includegraphics[height=30mm]{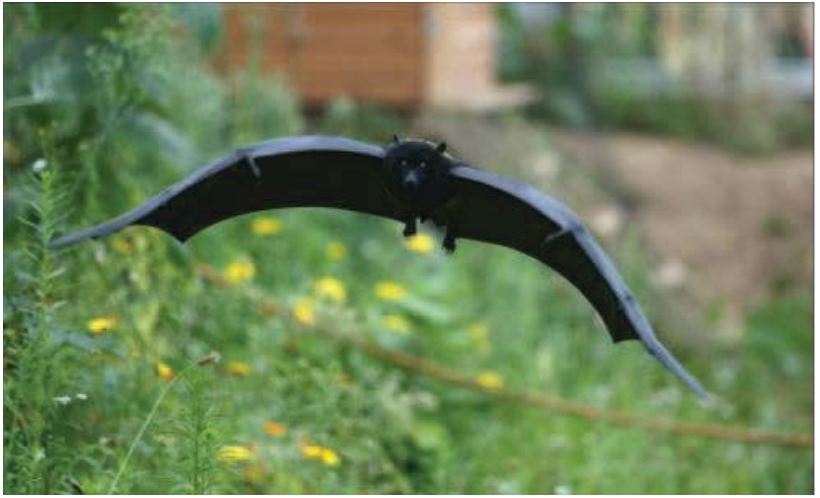}
\label{fig:camb_bend}
}
\quad
\subfigure[An arc is used to model wing shape when the bending or cambering happens.]{
\includegraphics[height=32mm]{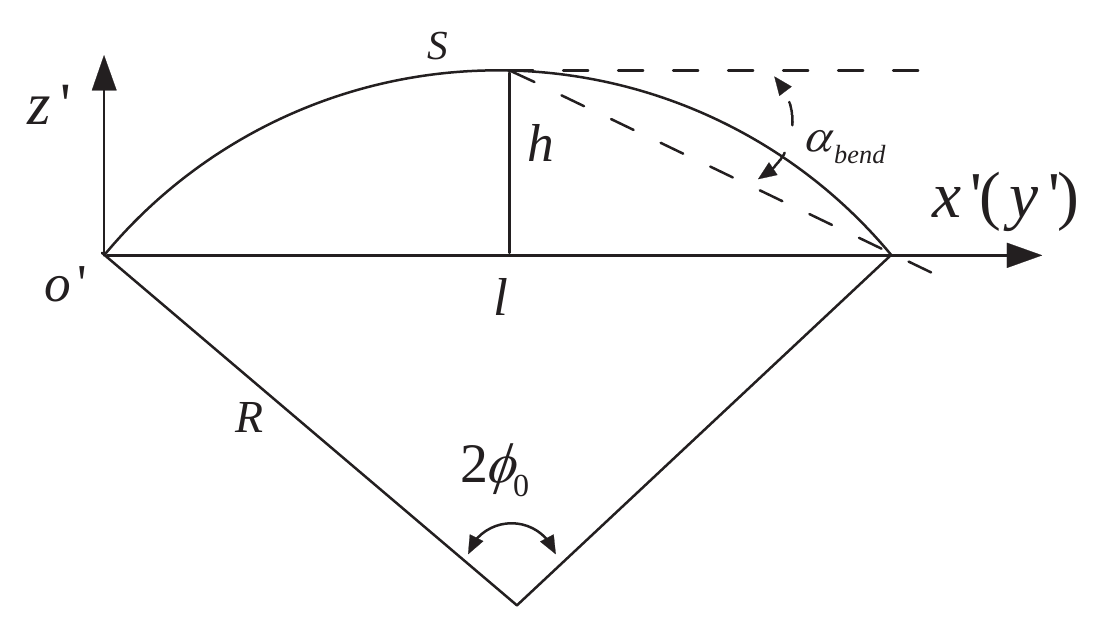}
\label{fig:arc}}
}
\caption{The wing shape and its arc-shaped model}
\label{fig:photo_arc}
\end{figure}

The bending deformation of a bat wing is defined as the morphing along the spanwise direction (see the photograph shown in \textbf{Fig.} \ref{fig:camb_bend}). Based on the observation in front view, a bat bends its wings more during upstroke than downstroke. The bending motion, modelled as an arc, is showed in \textbf{Fig.} \ref{fig:arc}. For an arc, there are some characteristic variables to determine the final shape. If the arc length $S$ is given, the chord length $l$ can be used to determine the shape of the arc. Then, the radius $R$, the central angle $2 \phi_0$, even the arch rise $h$ can be deduced.

\begin{figure}[h]
\centering
\begin{minipage}[t]{62mm}
\includegraphics[height=30mm]{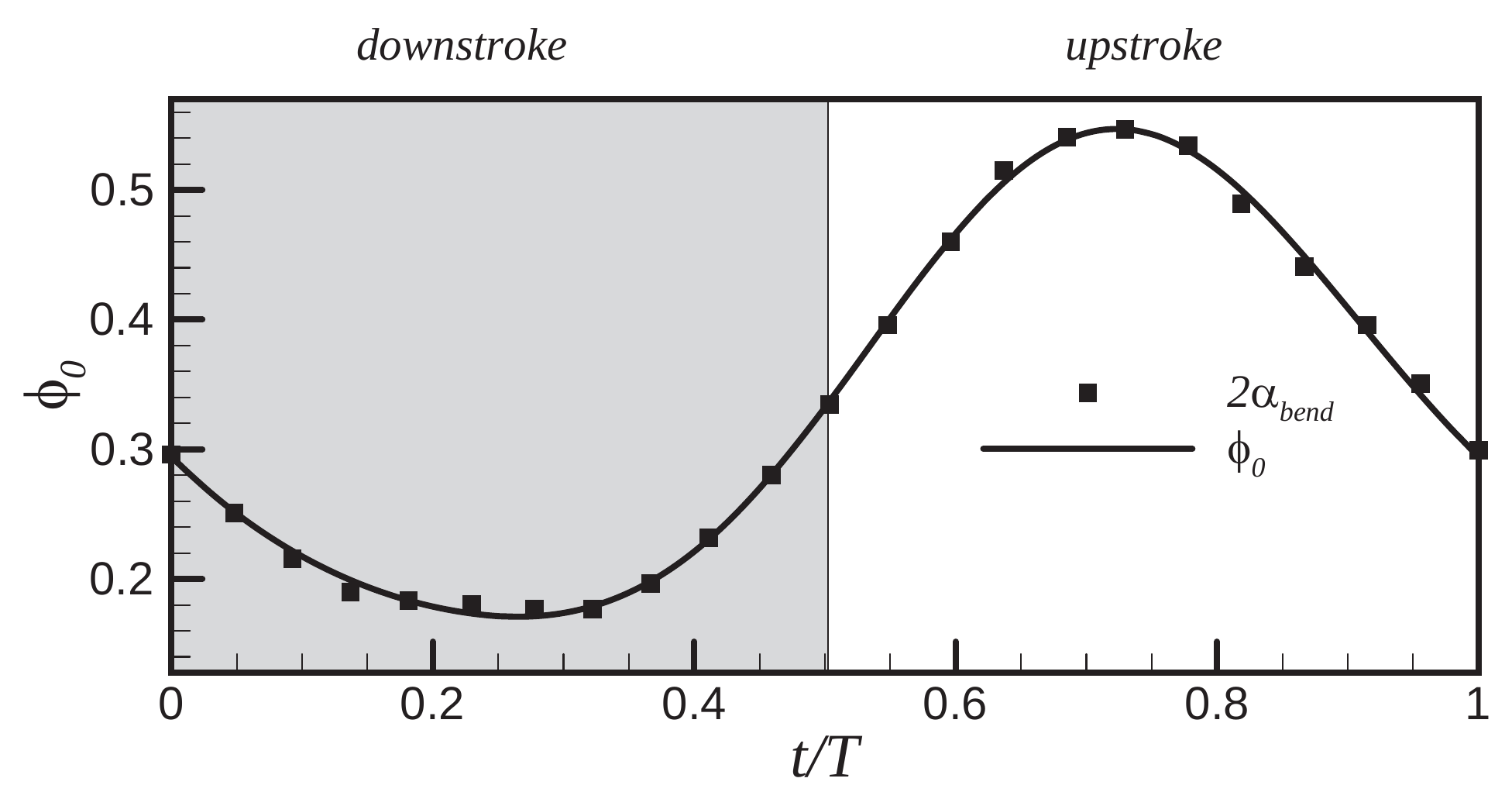}
\caption{The tendency of $\phi_0$ and $2\alpha_{bend}$ varying with the time, where $\alpha_{bend}$ is obtained from Busse\supercite{9}, $\phi_0$ is fitted with sinusoidal curve, and $T$ is a flapping cycle.}
\label{fig:bend_phi}
\end{minipage}
\qquad
\begin{minipage}[t]{62mm}
\includegraphics[width=62mm]{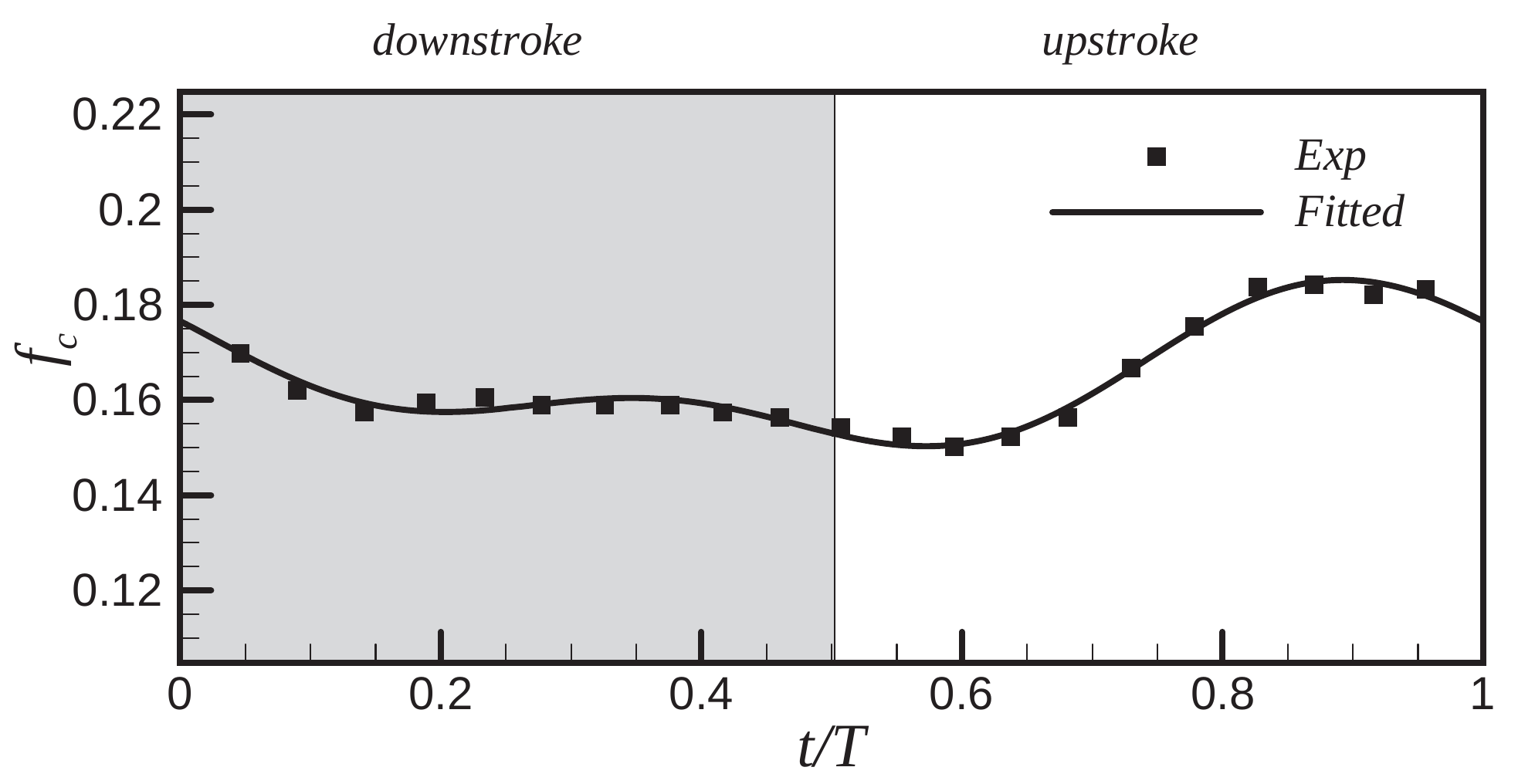}
\figcaption{The tendency of $f_c$ varying with the time, where ``\emph{Exp}'' indicates the experimental data from Busse\supercite{9}, while ``\emph{Fitted}'' is fitted by using a second order sinusoidal curve.}
\label{fig:fc}
\end{minipage}
\end{figure}

For the bending model, assuming the arc length $S$ to be a constant when the stretching and retracting wingspan is ignored (the variation of $S$ will be involved in the wing area-changing model), the changing central angle $2 \phi_0$ can be used to describe the bending motion, where $\phi_0$ is selected because it is twice as large as the angle of osculation $\alpha_{bend}$ which is measured in the work published by Busse et al.\supercite{9}. Furthermore, the bending motion was assumed to be synchronous for any spanwise section. So, the bending model is described as
\begin{equation}
\phi_0(t)=\overline{\phi}_0-\phi_{A} \sin(\omega t+ \varphi_{b})
\label{eq:phi_t}
\end{equation}
where $\overline{\phi}_0$ is the average value of $\phi_0$ during a flapping cycle, $\phi_A$ and $\varphi_{b}$ are the amplitude and the phase of the bending, respectively. According to the experimental data\supercite{9}, $\overline{\phi}_0$, $\phi_A$ and $\varphi_{b}$ are 0.34, 0.19 and 0.10, respectively. \textbf{Fig}. \ref{fig:bend_phi} shows the tendency of $\phi_0$ varying with the time, which is consistent with that of $2 \alpha_{bend}$ during the whole flapping cycle. So the simple harmonic model \eqref{eq:phi_t} depicts the main features of the bending deformation during the bat flapping flight.

\subsubsection{Cambering model}
The cambering deformation of a bat wing is defined as the morphing along the chordwise direction. Similar to the bending model, an arc-shaped deformation is assumed when the cambering deformation happens. Usually, the camber is determined with the chord length $l$ and the arch rise $h$, as showed in \textbf{Fig.} \ref{fig:arc}. These two variables can be directly measured in the experiments. When the arc length $S$ is given, a nondimensionalized arch rise is used to depict the camber, i.e.
\begin{equation}
f_c = h/l.
\label{eq:camber_fc}
\end{equation}
where subscript `$c$' denotes `camber'. For the bat wings, $f_c$ can be obtained from the camber of the fifth digit section, which has been measured by Busse et al.\supercite{9}. Similar to the bending model, the cambering motion is assumed to be synchronous for any chordwise section (spatial distribution). Because the cambering is more complex than the bending, so the second order term must be introduced into the model, i.e.
\begin{equation}
f_c(t)=A_c + A_{c1} \sin(\omega t+\varphi_{c1})+A_{c2} \sin(2\omega t + \varphi_{c2})
\label{eq:f_c}
\end{equation}
where the values of $A_c$, $A_{c1}$, $A_{c2}$, $\varphi_{c1}$ and $\varphi_{c2}$ are based on the experimental data fitting. As shown in \textbf{Fig}. \ref{fig:fc}, the tendency of $f_c$ varying with the time is reasonable when the parameters in equation \eqref{eq:f_c} are set in \textbf{Tab}. \ref{tab:fc}.

\begin{center}
\abovecaptionskip 0pt \belowcaptionskip 1pt
\renewcommand{\arraystretch}{1.2}
{
\scriptsize \tabcaption{The parameters in the cambering model.}\label{tab:fc}
\noindent
\begin{tabular*}{\textwidth}{@{\extracolsep{\fill}}@{~~}ccccccc}
\toprule
	{$A_c$}	&{$A_{c1}$} &{$A_{c2}$} &{$\varphi_{c1}$} &{$\varphi_{c2}$}\\
    \midrule
    0.16 		& 0.013 	& -0.008	& 2.02    & 0\\
\bottomrule
\end{tabular*}
\small
}\\[4mm]
\end{center}

\subsubsection{Area-changing model}
Besides of the twisting, bending and cambering deformation, the surface area of a bat wing is changing rhythmically with the flapping. In fact, the wing area-changing is under the control of the upper limb. On the one hand, it is found that the wing span during the upstroke is less compared to that during the downstroke. On the other hand, the length along the chordwise direction changes little during the whole flapping process. So, the area of the wing membrane increases (or decreases) when the wingspan is stretched (or retracted) by the upper limb.

The span ratio ($SR$), defined as the span at mid-upstroke divided by the span at mid-downstroke, was measured in the experiments\supercite{9}. Its value reaches a level around $0.7\pm0.03$ at intermediate speed (around 3 m/s). That means, the change of wingspan is so large that it cannot be ignored in bat flight. Therefore, we suppose the area of model-wing changes sinusoidally, i.e. the wingspan (arc length $S$) is changing with time,
\begin{equation}
S(t)=S_0 (1 + A_s \sin \omega t)
\label{eq:area_changing}
\end{equation}
where $S_0$ is the initial wingspan, $A_s$ is the nondimensional amplitude of the changing wingspan. So, the span ratio is $SR = (1-A_s)/(1+A_s)$. When $SR=0.72$ at the speed of 3 m/s, we obtain $A_s \approx 0.16$.

\section{Results and Discussions}
A plate of aspect ratio 3 is used to model a bat wing and a three dimensional unsteady panel method has been developed to predict the aerodynamic forces generated by the flapping model-wing with leading edge separation. Details of the method were described in our previous paper\supercite{3}. The lift and drag coefficients are defined as $C_L= F_z/0.5\rho U^2 A$ and $C_D= F_x/0.5\rho U^2 A$, respectively, where $F_z$ is the vertical upward force (lift), $F_x$ is the horizontal backward force (drag), $\rho$ is the air density, $U$ is the inflow velocity and $A$ is the reference plate area. For the present rectangular wing-model, the reference plate area is defined as the product of the initial wingspan and the chord length, and its nondimensional value is 3 in the present problem.

\subsection{Aerodynamic performance}
In this subsection, the influence to the aerodynamic forces will be presented one by one according to the elemental morphing models, the bending model, the cambering model and the area-changing model. In order to eliminate the influence of the twisting that has been studied in the previous work\supercite{3}, the twisting amplitude is set to zero, $\alpha_A = 0$. The resolution of the grid is $20 \times 30$, shown in \textbf{Fig}. \ref{fig:grid}.

\subsubsection{Effects of bending}
As mentioned in \S 2.2.1, the bending model is governed by equation \eqref{eq:phi_t} and there are three parameters controlling the motion, which are the basic bend $\overline{\phi}_0$, the bending amplitude $\phi_A$ and the phase $\varphi_b$. The aerodynamic forces generated by the flapping wing with the bending deformation will be compared with that by the purely flapping wing. Then the influence of each parameter to the forces is discussed.

\begin{figure}[h]
\centering
\begin{minipage}[t]{62mm}
\includegraphics[width=65mm]{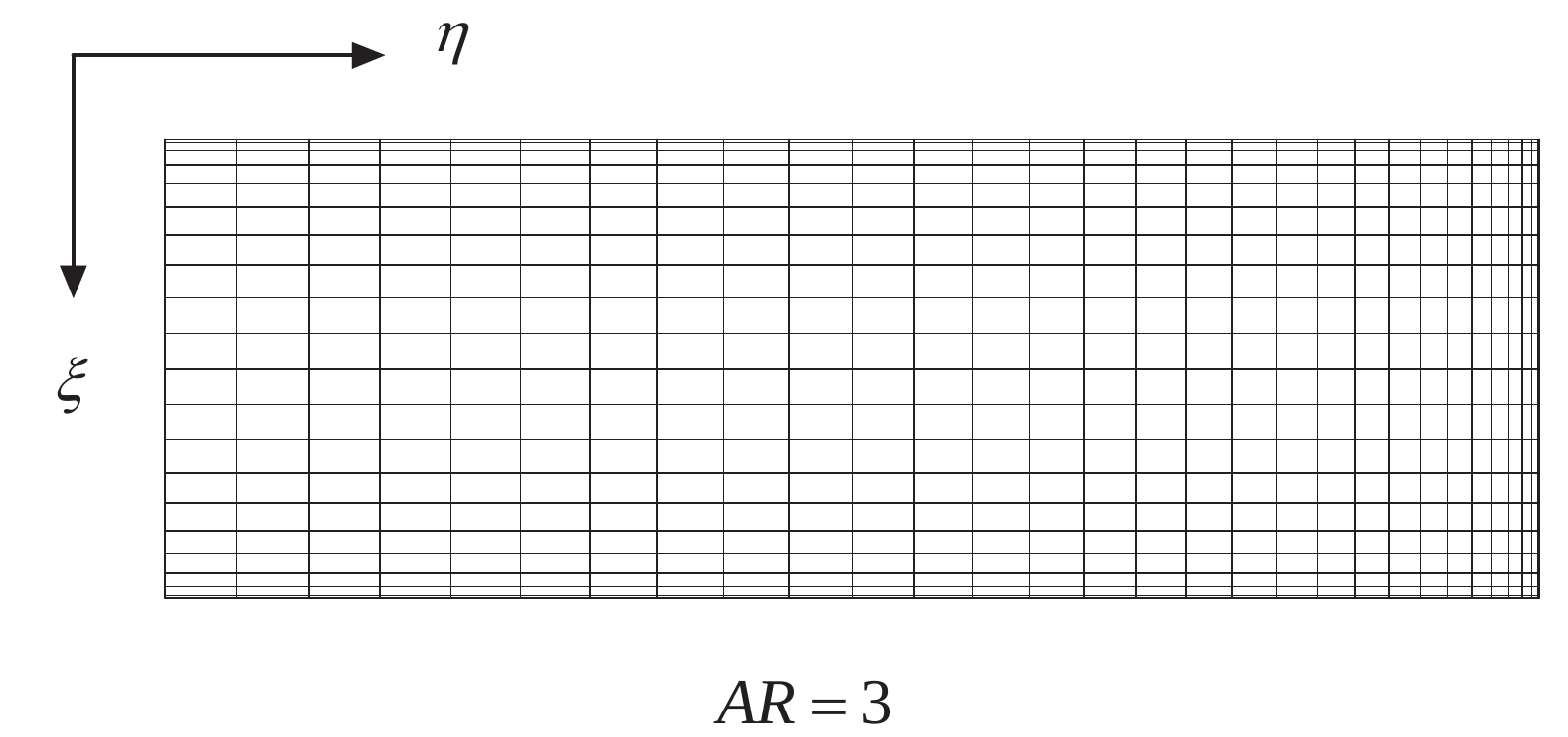}
\figcaption{The grid of a model wing whose resolution is $20\times 30$, with 20 in the chordwise direction and 30 in the spanwise direction.}
\label{fig:grid}
\end{minipage}
\qquad
\begin{minipage}[t]{62mm}
\includegraphics[width=60mm]{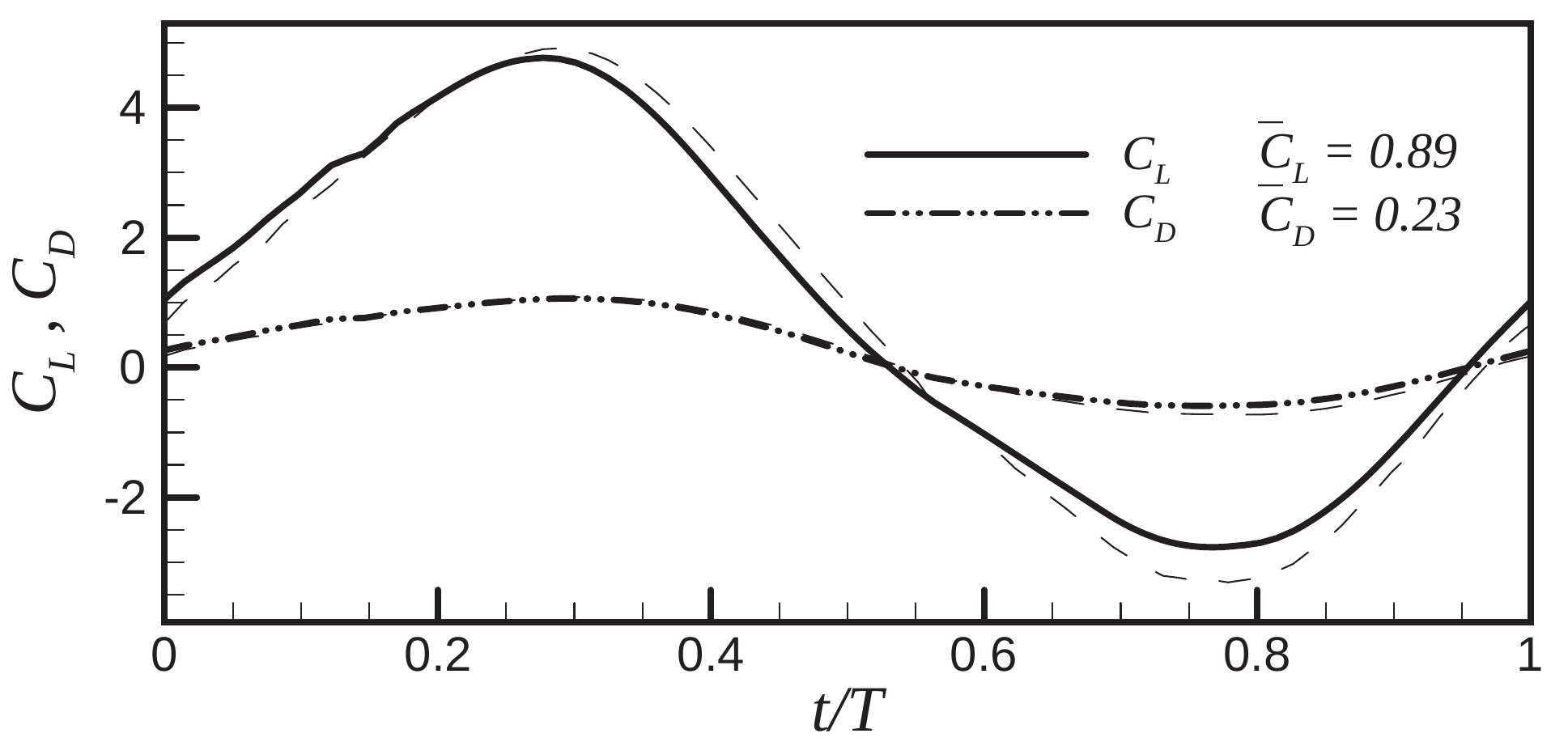}
\figcaption{The time-dependent lift and drag coefficients of a bending flapping wing. The dashed lines correspond to the forces of a purely flapping wing.}
\label{fig:bending_force}
\end{minipage}
\end{figure}

First, as shown in \textbf{Fig}. \ref{fig:bending_force}, the lift and drag curves are presented. It seems that the difference between the forces produced by the bending wing and the purely flapping wing is small. Furthermore, the time-averaged lift $\overline{C}_L$ (0.89) is larger than that of the flapping rigid-wing (0.79). The increment of $\overline{C}_L$ is about $12.6\%$, while there is no difference for the time-averaged drag $\overline{C}_D$, both of which are 0.23.

\begin{figure}[h]
\centering
\mbox{
 \subfigure[The lift generated by the bending wing]{
 \includegraphics[height=35mm]{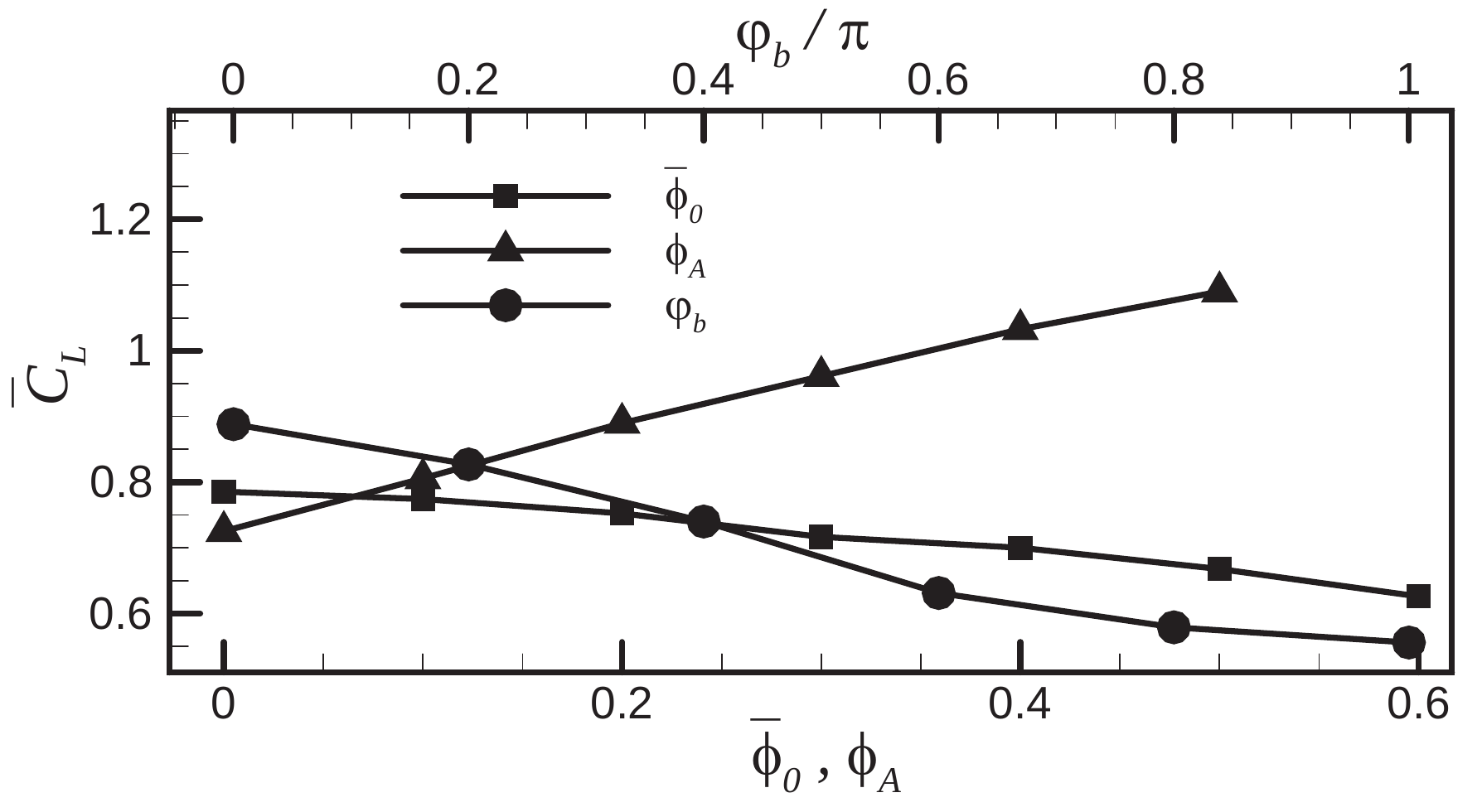}
 \label{fig:bending_cl}
 }
 \quad
 \subfigure[The drag generated by the bending wing]{
 \includegraphics[height=35mm]{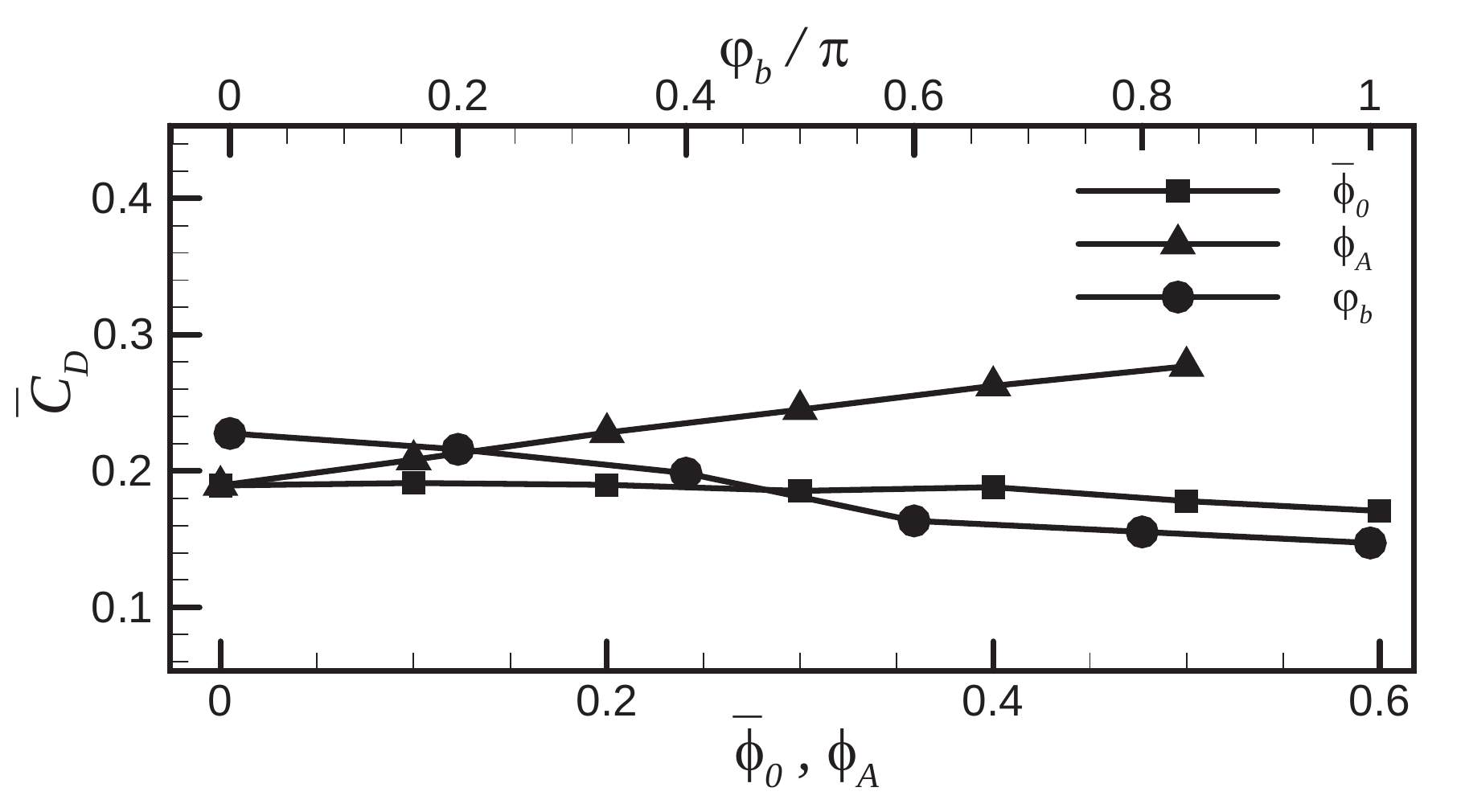}
 \label{fig:bending_cd}
 }
}
\figcaption{The effects of the bending deformation.}
\label{fig:effect_bending}
\end{figure}

Second, the effects of three parameters to the forces are shown in \textbf{Fig}. \ref{fig:effect_bending}. It is found that the basic bend $\overline{\phi}_0$, the bending amplitude $\phi_A$ or the phase $\varphi_b$, changes the averaged drag little (see \textbf{Fig}. \ref{fig:bending_cd}). But the larger bending amplitude, the larger averaged lift is, while the larger basic bend or the phase, the less lift is (see \textbf{Fig}. \ref{fig:bending_cl}).

So, it becomes known that the appropriate dynamic deformation of bend along the spanwise direction can enhance the lift with the little influence to the drag. To increase the bending amplitude is an efficient way to enhance the lift.

\subsubsection{Effects of cambering}
\begin{figure}[h]
\centering
\mbox{
  \subfigure[The time-dependent lift and drag caused by the cambering wing during a stroke]{
  \includegraphics[width=60mm]{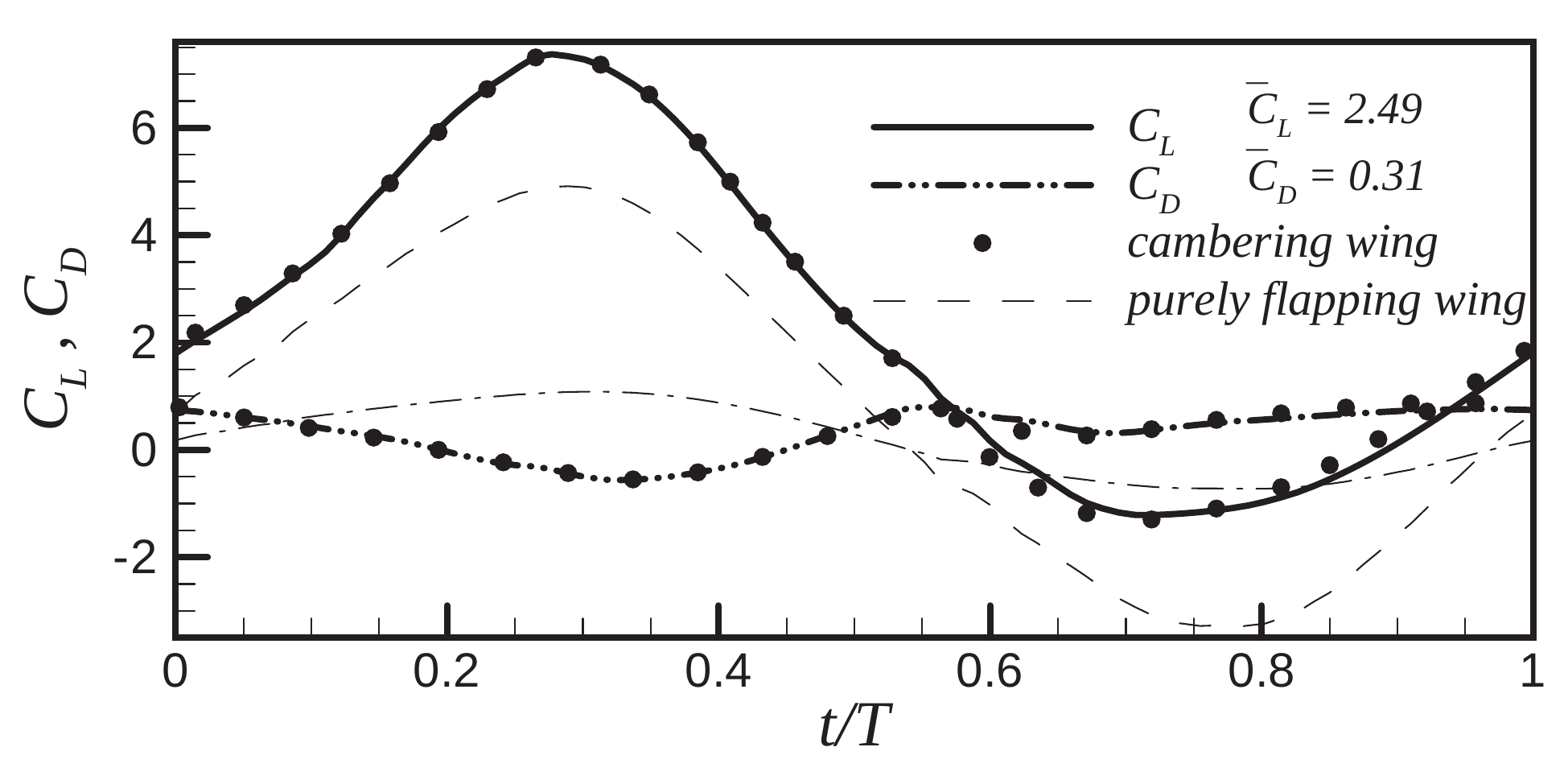}
  \label{fig:cam_force_time} }
  \quad
  \subfigure[The mean lift and drag varying with the basic camber ${A}_c$]{
  \includegraphics[height=30mm]{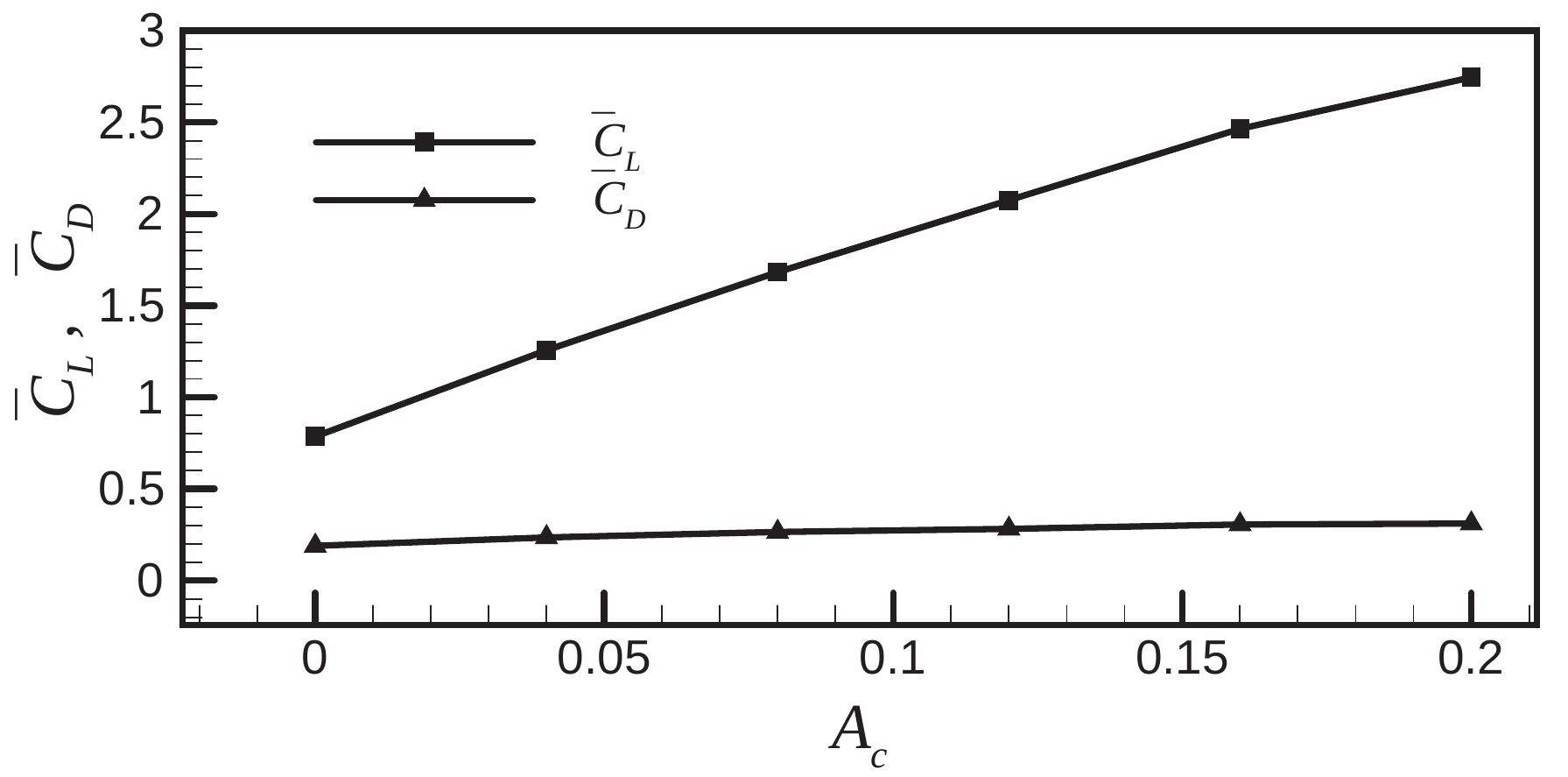}
  \label{fig:cam_bar}}
}
\figcaption{The effects of the cambering motion where the controlling parameters are listed in \textbf{Tab}. \ref{tab:fc}.}
\label{fig:cam_force}
\end{figure}

In the cambering model, the amplitudes of the harmonic terms, $A_{c1}$ and $A_{c2}$, are less than one-tenth of the basic camber $A_c$, as shown in \textbf{Tab}. \ref{tab:fc}. So we compared the aerodynamic forces in three cases shown in \textbf{Fig}. \ref{fig:cam_force_time}, the purely flapping, the flapping with the basic camber and that with the cambering deformation. It is indicated that there is almost no difference for the forces between the latter two cases. Relative to the purely flapping wing, the flapping wing with camber or cambering deformation can elevate the lift during the whole stroke. And it changes the shape of the drag curve not only in the downstroke but also in the upstroke. Furthermore, the effect of the basic camber deformation ($A_c$) is discussed with the fixed cambering amplitudes ($A_{c1}=0.013$ and $A_{c2}=-0.008$). According to the time-averaged lift and drag shown in \textbf{Fig}. \ref{fig:cam_bar}, it is found that the basic camber deformation enhances the lift greatly but changes the drag little. When the basic camber grows to 0.16, the lift rises to 2.49, three times as much as that of the purely flapping wing.

On the view of aerodynamics, the camber is indispensable to produce the high lift. In the published experimental data about two bats\supercite{17}, the wing loads are $13.4 \text{N/m}^2$ and $14.7 \text{N/m}^2$, respectively. The mean lift coefficients required to stay aloft at speed of $3\text{m/s}$ are $2.48$ and $2.72$, where the air density is $\rho = 1.20\text{kg/m}^3$. The predicted average lift is sufficient to support the weight of the bats when the cambering deformation occurred.

\begin{figure}[h]
\centering
\mbox{
  \subfigure[The lift curves]{
  \includegraphics[height=30mm]{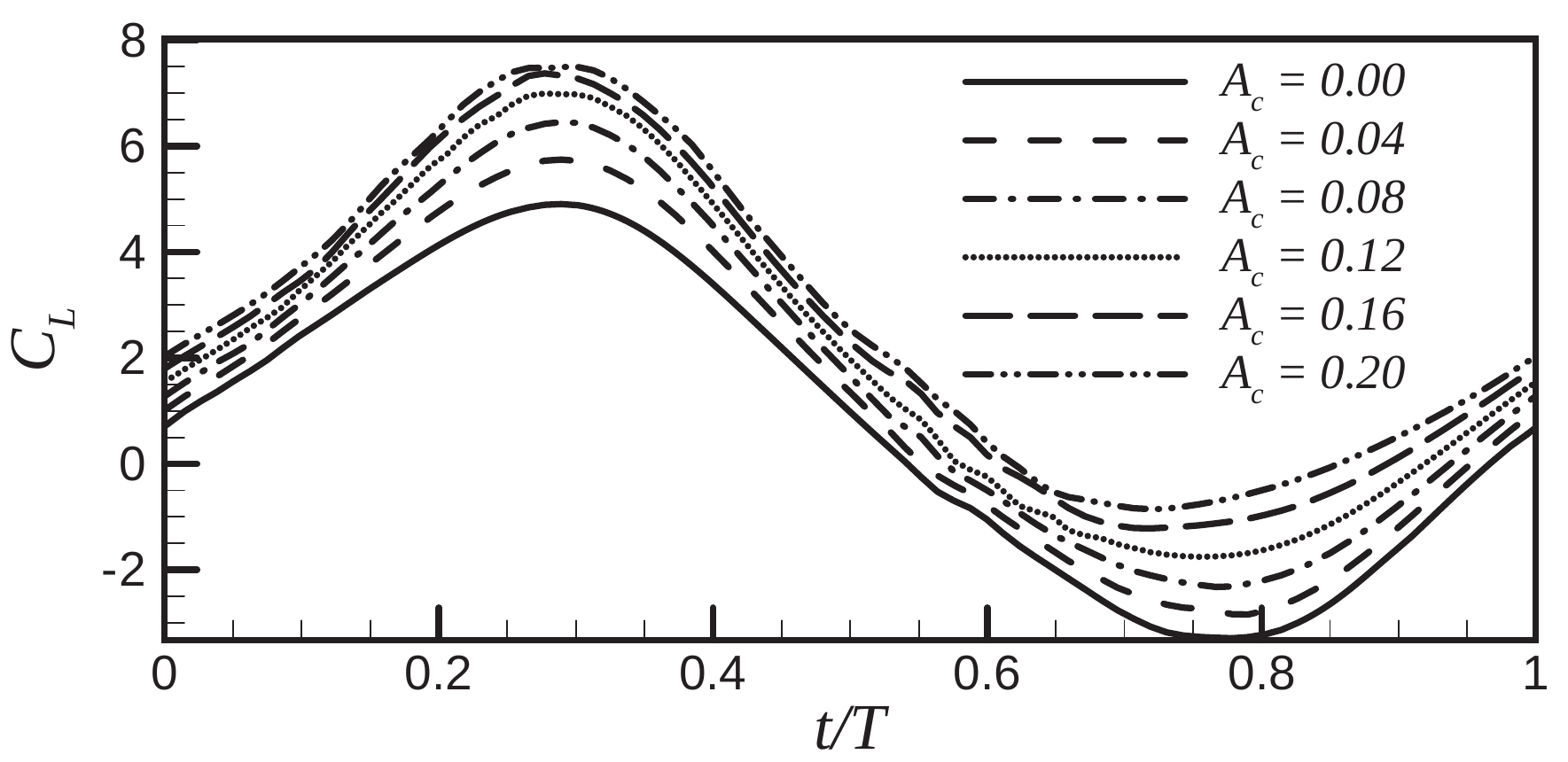}
  \label{fig:cam_CL} }
  \quad
  \subfigure[The drag Curves]{
  \includegraphics[height=30mm]{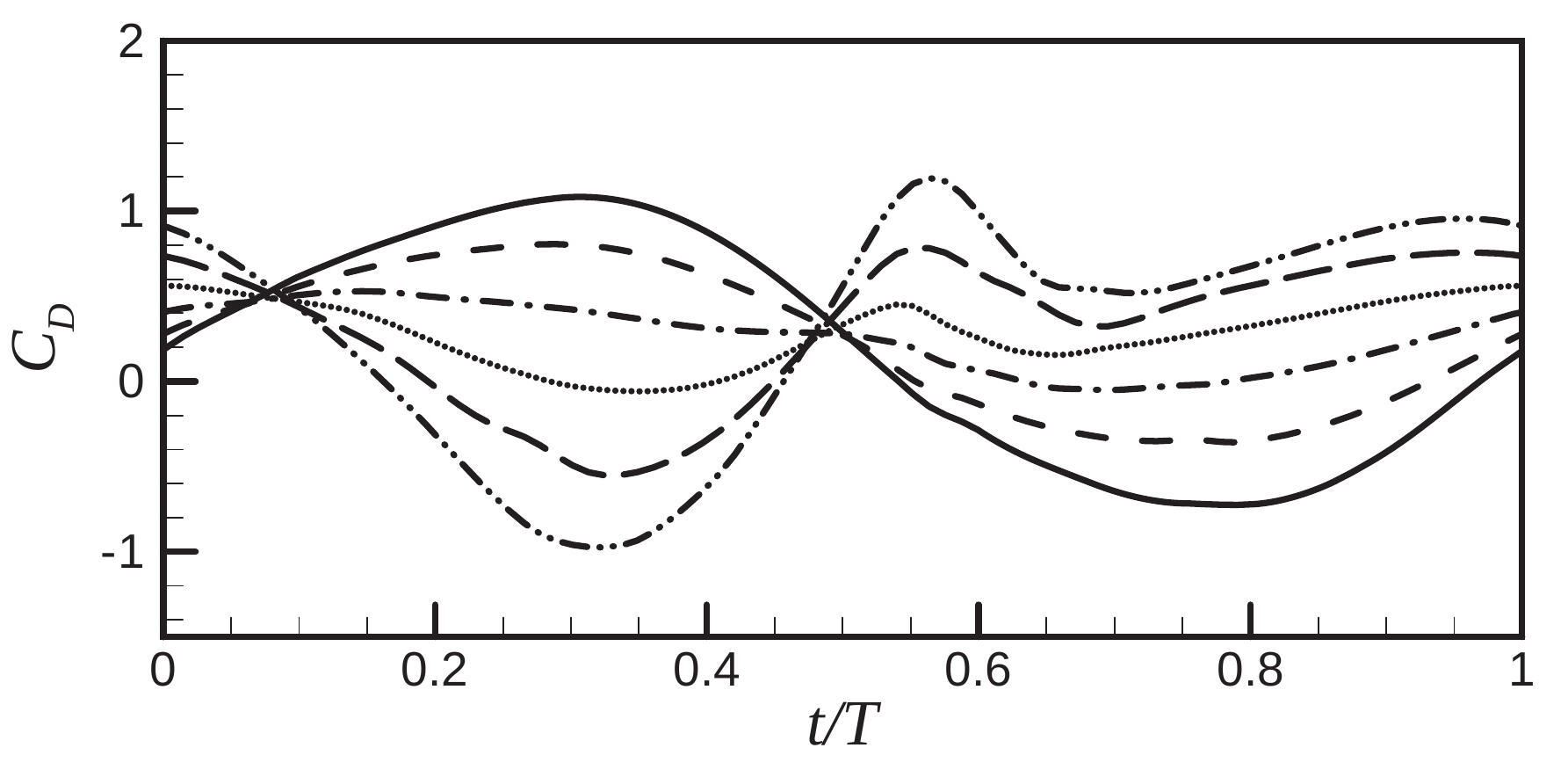}
  \label{fig:cam_CD}
  }
}
\figcaption{The variation of the aerodynamic forces generated by the different cambered wings (${A}_c$)}
\label{fig:cam_aero_force}
\end{figure}

Moreover, the details of the lift and drag curves with different basic camber are presented in \textbf{Fig}. \ref{fig:cam_aero_force}. For the lift, the bigger the basic camber is, the higher the curve is elevated (shown in \textbf{Fig}. \ref{fig:cam_CL}). For the drag, the variation tendency changes with $A_c$ (shown in \textbf{Fig}. \ref{fig:cam_CD}). When the basic camber $A_c$ is less than $0.08$, the curves is almost of the same shape, but when $A_c > 0.08$, the minus drag happens during the downstroke which is positive for the bat in forward flight. So, the higher lift and lower drag are improved in the duration of downstroke. This phenomenon will help to optimize the aerodynamic forces during the downstroke.

\subsubsection{Effects of area-changing}
\begin{figure}[h]
\centering
\mbox{
  \subfigure[The lift/drag of a flapping wing with area-changing. The dashed lines correspond to the forces of a purely flapping wing]{
  \includegraphics[height=30mm]{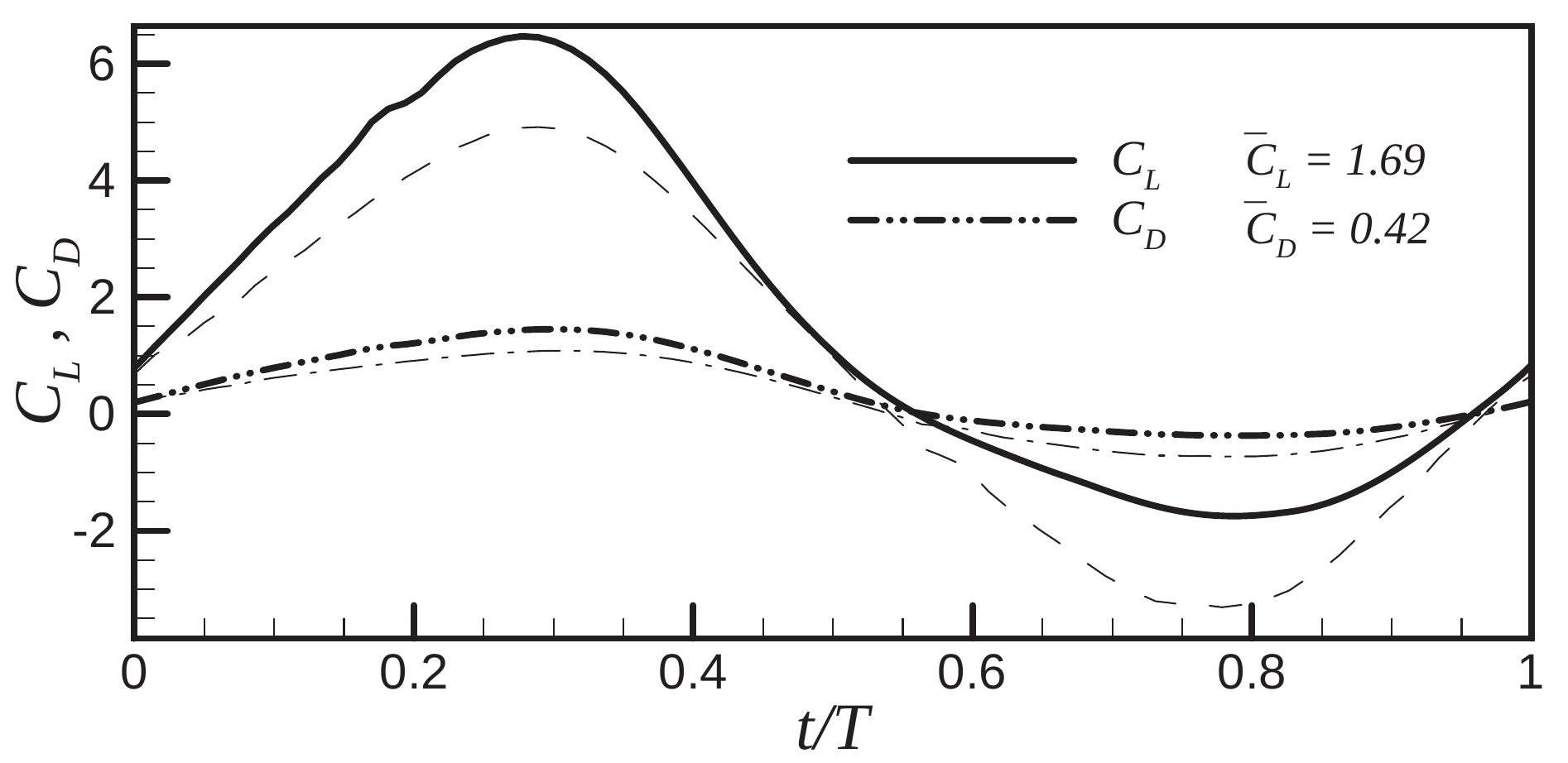}
  \label{fig:area_force_time} }
  \quad
  \subfigure[The mean lift and drag varying with the area-changing amplitude $A_s$.]{
  \includegraphics[height=30mm]{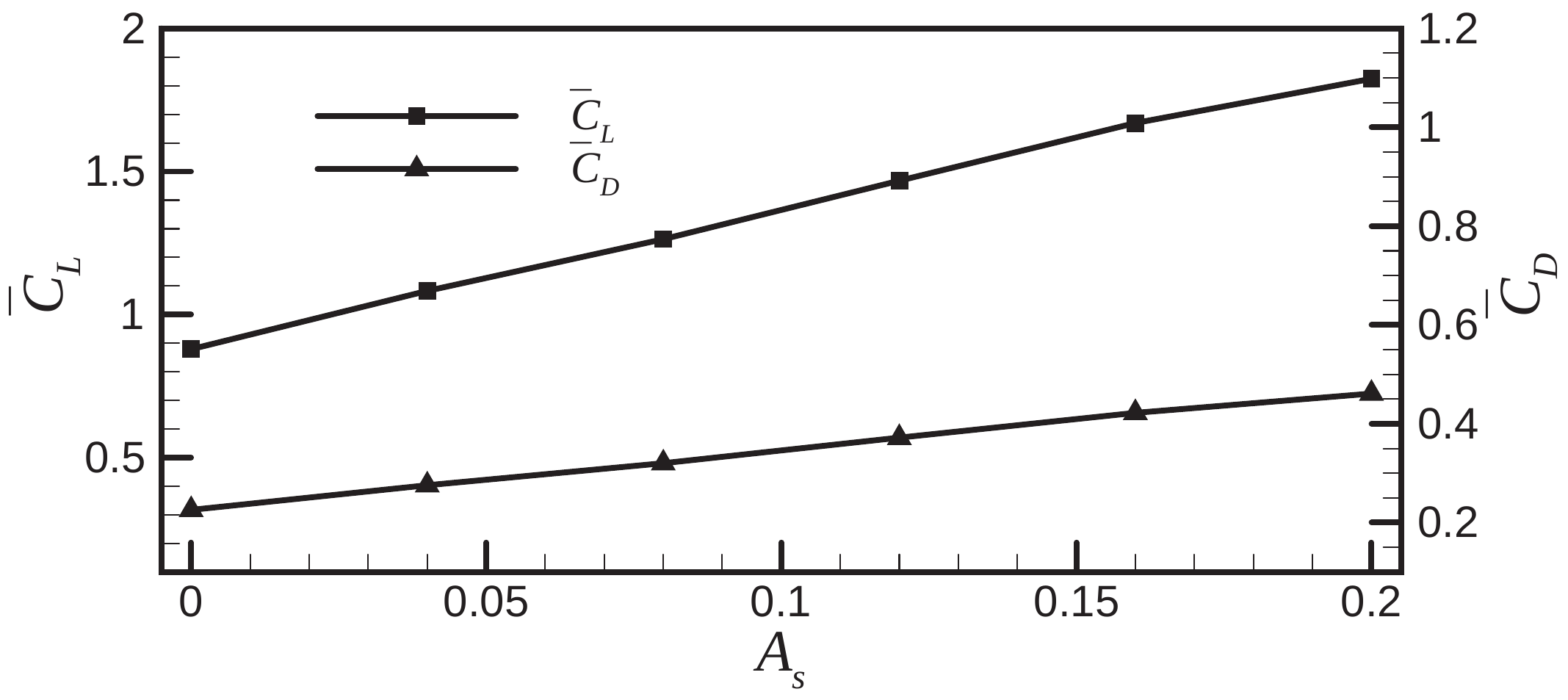}
  \label{fig:area_bar} }
}
\figcaption{The effects of the area-changing}
\label{fig:aero_forces}
\end{figure}

According to \textbf{Eq}. \eqref{eq:area_changing}, the wing surface area depends on the wingspan, which is changing with time. In this subsection, the effects of area-changing are investigated. First, a typical case is shown in \textbf{Fig}. \ref{fig:area_force_time} ($A_s=0.16$), which indicates that the variation tendency of $C_L(t)$ and $C_D(t)$ curves of the area-changing model is as same as that of the purely flapping wing. But the peak of the lift becomes higher during the downstroke and the trough (minus lift) becomes smaller during the upstroke. It is a good phenomenon to enhance the time-averaged lift during the whole stroke. The mean lift is $1.69$, a twofold increase compared with the purely flapping wing. The mean drag is also increased by the elementary area-changing model, which rises up to $0.42$.

Second, the mean lift and drag varying with the different values of the nondimensional amplitude of the changing wingspan are shown in \textbf{Fig}. \ref{fig:area_bar}. It is found that, the mean lift increases linearly with $A_s$, and the mean drag increases too but its slope is smaller than that of the lift. So the lift-drag ratio grows with the amplitude of the changing wingspan although it grows slightly. As same as the bending and cambering models, the area-changing model can enhance the lift and benefit the bat flight at the moderate or slow speed.

\subsection{Mechanisms for the aerodynamic force enhancement}
The effects on the aerodynamic forces of three types of morphing models have been investigated and discussed in the last substion. The cambering model has a great positive impact on the lift, followed by the area-changing model, and then the bending model. As discussed in the previous paper\supercite{3}, the mechanism of the twisting deformation for generating the thrust is that of the supination/pronation motion in insect flight. Now, the mechanisms of the elementary morphing (bending, cambering and area-changing) for the aerodynamic force enhancement will be discussed in this subsection.

\subsubsection{Added mass effects and vortex effects}
\begin{center}
\abovecaptionskip 0pt \belowcaptionskip 1pt
\renewcommand{\arraystretch}{1.2}
{\scriptsize \tabcaption{The added mass effects, vortex shedding effects and total lift.}\label{tab:cam_mechan_add_vortx}
\noindent\begin{tabular*}{\textwidth}{@{\extracolsep{\fill}}@{~~}ccccccc}
\toprule%
    &\multicolumn{2}{c}{total force}
    &\multicolumn{2}{c}{added mass effects}
    &\multicolumn{2}{c}{vortex effects}\\
    \midrule
		{$\overline{C}_L$}&downstroke &upstroke	 &downstroke &upstroke &downstroke &upstroke\\
    \midrule
    Purely flapping wing &3.22 		& -1.64 	& 0.06	& -0.06    & 3.16  &-1.58\\
    Twisting wing &2.74    &-0.78      &0.12   &-0.12     & 2.62  &-0.66\\
    Bending wing   &3.12        &-1.35      &0.014  &-0.014    & 3.11  &-1.34\\
    Cambering wing &4.94 		& -0.004	& 0.06	& -0.06	   & 4.88  &0.056\\
    Area-changing wing &4.08    &-0.75      &0.12   &-0.12     & 3.96  &-0.63\\
\bottomrule
\end{tabular*}
\small }\\[4mm]
\end{center}

In the present paper, the unsteady panel method is used to predict the aerodynamic forces acting on the morphing model-wing in bat flight. Similar to the theoretical modelling in analyzing insect flight\supercite{16, 17}, one of the advantages of this method is to separate the force into two parts: the added mass effects and the vortex effects. The added mass effect is an instantaneous force acting on the accelerating body, and the vortex effect depends on the vortex shedding from the body which includes historical effects.

In \textbf{Table} \ref{tab:cam_mechan_add_vortx}, the added mass effects, vortex effects and total lift during the downstroke and the upstroke are presented. It is found that, (A) the added mass effects are much small not only during the downstroke but also during the upstroke, and their integrals during the whole stroke are zero; (B) the much higher positive lifts during downstroke are generated than that during the upstroke for the morphing wings, especially for the cambering one; (C) compared with the purely flapping wing, the minus lifts during the upstroke produced by the morphing wings are reduced greatly, as is of benefit to the mean lift enhancement during the whole stroke.

Since the added mass effects cannot affect the mean lift in a cycle, the forces produced by the moving vortex are the unique effects for the morphing wings in bat flight. That means the vortex control is a key to generate high aerodynamic forces.

\subsubsection{Asymmetry of the cambered wing}

As shown in \textbf{Table} \ref{tab:cam_mechan_add_vortx}, it is a remarkable phenomenon that the cambering wing generates very high lift during the downstroke and minimizes the minus lift during the upstroke. Why does it affect the lift? In the steady aerodynamics, there is a mechanism to reveal the effect of a camber wing, i.e. the effective angle of attack of the camber wing is larger (or smaller) than the actual geometric angle of attack, which leads to the higher (or lower) lift of the wing in a uniform flow.

In Katz's book\supercite{18} (see equation 5.85), a thin airfoil with a parabolic camber is introduced and its corresponding lift coefficient is thus
\begin{equation*}
C_L = 2\pi (\alpha + 2 h/l)
\end{equation*}
where, $\alpha$ is the actual geometric angle of attack, and $h/l$ is the nondimensional arc rise which is defined as $f_c$ in the present paper. Actually, $\alpha + 2h/l$ is the effective angle of attack, and $2h/l$ is regarded as the added angle of attack. The equation indicates that there exists lift even if the actual geometric angle of attack is zero. This lift enhancement by effective angle of attack is called the camber effect. In fact, the camber effect depends on the asymmetry of the wing. On the one hand, the higher asymmetry, the bigger difference between the effective angle of attack and the actual geometric one is. On the other hand, the effective angle of attack is increased by the traditional camber ($h>0$) but decreased by the anti-camber ($h<0$). In other words, the lift caused by the cambered wing is different with that by the anti-cambered wing.

Although the flow around the cambering and flapping bat wing is unsteady and the actual geometric angle of attack is required to be small in the theoretical derivation, the qualitative analysis due to the steady conclusion can be used to reveal the secret of high lift generation. During the downstroke, the effective angle of attack is larger than the geometric one by $2h/l$, which results in the higher lift generation. But during the upstroke, the `anti-camber' makes the effective angle of attack smaller by $2h/l$, which results in the negative lift smaller. Therefore, the asymmetry of the cambered wing is the main mechanism to generate high lift in flapping flight.

\subsubsection{Amplifier for the aerodynamic forces}
Whatever the morphing motion is, the downstroke benefits the flight for the high lift generation, even the thrust is generated by the twisting wing\supercite{3}. And the upstroke with the negative lift is harmful while it also generates the thrust. A simple idea is to amplify the positive effects in the downstroke and to reduce the negative effects in the upstroke.

To change the surface area of wing is an effective way to help the flight. As mentioned in equation \ref{eq:area_changing}, the area of the wing varies sinusoidally and $A_s$ describes the amplitude of the wing surface area changing. During the downstroke, the area is larger than the reference area $S_0$ for the wingspan is stretched. At mid-downstroke, the surface area reaches the maximum value. During the downstroke, the area is smaller than $S_0$ for the wingspan is retracted. At mid-upstroke, it reaches the minimum. The large wing-area amplifies the high lift in the downstroke and the small wing-area reduces the minus lift in the upstroke. Therefore, the mechanism of the wing area-changing is the effect of an amplifier to amplify the positive lift and reduce the minus lift.

To some extent, the bending deformation is another amplifier in bat flight. As shown in figure 3(b), the wingspan bends less during the downstroke than the upstroke, which means the projected area is larger in the downstroke. So, it is easy to understand the effects of bending.

\begin{figure}[h]
\centering
\includegraphics[width=60mm]{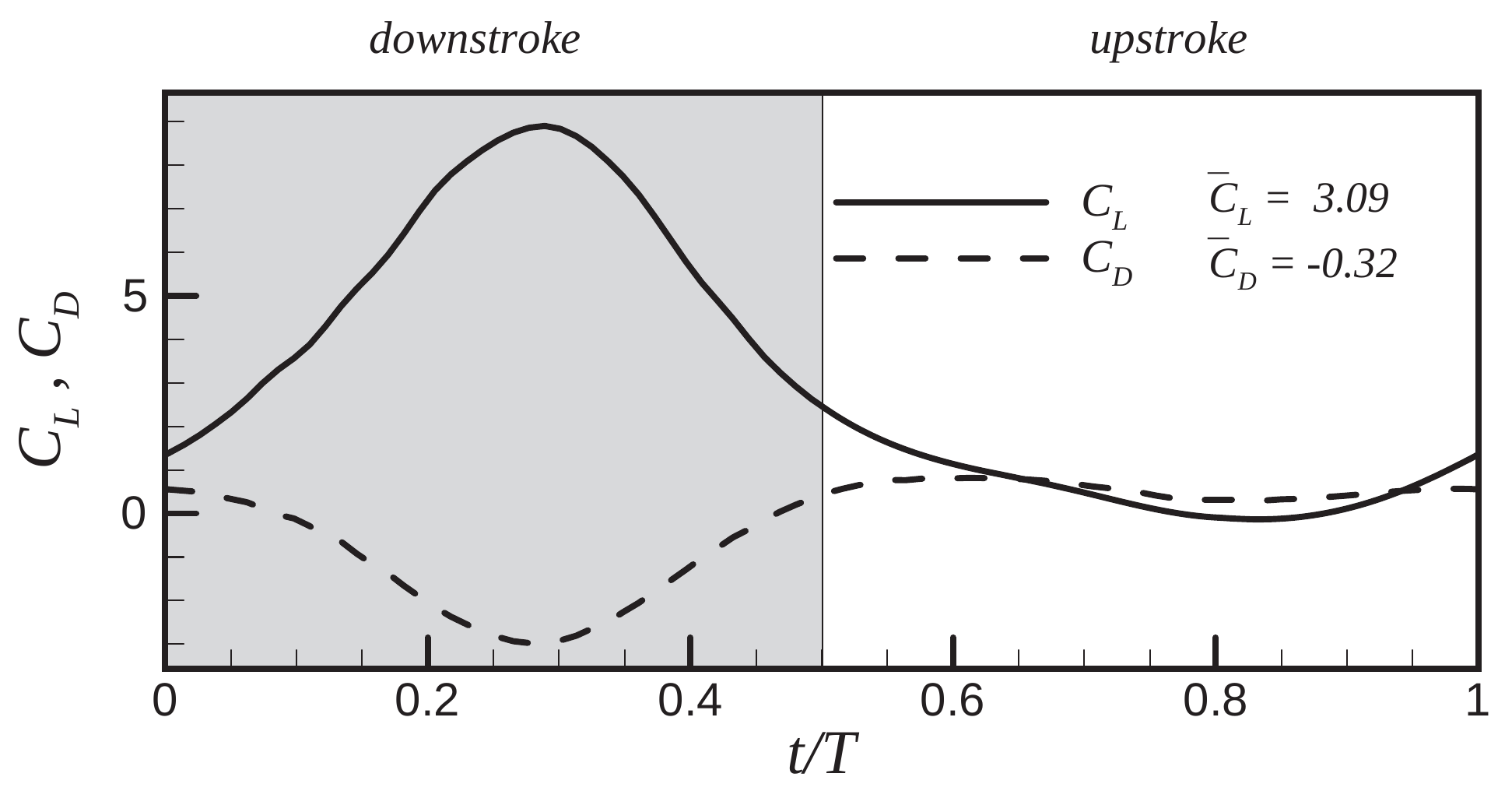}
\figcaption{The forces generated by the integrated morphing model-wing which contains twisting ($\alpha_A=28\,^{\circ}$), bending, cambering and area-changing models}
\label{fig:integrated_model}
\end{figure}

Furthermore, the forces generated by the integrated morphing model-wing (containing twisting, bending, cambering and area-changing) are shown in \textbf{Fig}. \ref{fig:integrated_model}, where the twisting amplitude is $\alpha_A=28\,^{\circ}$. The mean lift is $3.09$ and the mean thrust is $0.32$. It is worth noting that the twisting wing can generate the thrust not only during the downstroke but also during the upstroke while the other elementary morphing wings almost generate the drag during the upstroke. So the small drag (or thrust) is generated by integrated morphing wing during the upstroke, even the small lift. It is concluded that the lift and thrust are mainly generated during downstroke, which benefit the amplifiers running.

Therefore, the effect of amplifier is the mechanism of the wing area-changing or bending for the aerodynamic forces enhancement.

\section{Conclusions}

As a unique flying mammal, bat is able to manipulate their fingers into controlling the flexible wing-membrane, so the aerodynamic forces required in flight can be produced by the large actively morphing wings. Followed the twisting model in the previous study\supercite{3}, three elementary morphing models (the bending model, cambering model and area-changing model) are proposed in this paper and the aerodynamic forces generated by the model-wing are investigated by an unsteady panel method. The results indicate that compared with the purely flapping wing whose mean lift coefficient is 0.79, the cambering wing can generate much high lift ($\overline{C}_L=2.49$) in flapping flight, the area-changing wing can produce high lift also ($\overline{C}_L=1.69$), and the bending wing can enhance lift ($\overline{C}_L=0.89$).

Furthermore, the mechanisms for the aerodynamic force enhancement are discussed. First, because of the no effect of the added mass during a flapping cycle, the vortex control is a main approach to generate the mean high aerodynamic forces. Second, the asymmetry of the cambered wing is of great benefit to high lift generation. Last, the amplifier effect for the aerodynamic forces is the mechanism of the wing area-changing or bending and it amplify the forces during the downstroke and reduce them during the upstroke. It is also found that the lift and thrust are mainly generated during the downstroke while almost no forces during the upstroke by the integrated morphing model-wing, which includes all the mechanisms for the aerodynamic forces enhancement.
\\
\\
\noindent  \textbf{Acknowledgements}\quad\footnotesize
The authors are most grateful to Professor TONG Bing-Gang and Associate Professor BAO Lin for their valuable advices.   
\end{CJK*}
\end{document}